\documentclass[aps,prl,twocolumn,showpacs]{revtex4}
\usepackage{bbm,amsmath,amssymb,amsbsy,graphicx,txfonts,subfigure,epstopdf}

\usepackage[latin1]{inputenc}
\newcommand{\mathsym}[1]{{}}

\newcommand{\id}{\mathbbm{1}}

\newcommand{\gr}[1]{\boldsymbol{#1}}
\newcommand{\ket}[1]{|#1\rangle}
\newcommand{\bra}[1]{\langle#1|}

\newcommand{\sig}{{\bf V}}
\newcommand{\diag}{\text{diag}}
\newcommand{\eq}[1]{Eq.~(\ref{#1})}

\begin{document}
\title{Controllable Gaussian-qubit interface for extremal quantum state engineering}
\author{G. Adesso$^1$, S. Campbell$^2$, F. Illuminati$^{3}$, and  M. Paternostro$^2$}
\affiliation{$^1$School of Mathematical Sciences, University of Nottingham, University Park, Nottingham NG7 2RD, United Kingdom\\
$^2$School of Mathematics and Physics, Queen's University, Belfast BT7 1NN, United Kingdom\\
$^3${Dipartimento di Matematica e Informatica, Universit\`a degli Studi di
Salerno, CNR-SPIN, CNISM, Unit\`a di Salerno, and INFN, Sezione di Napoli -
Gruppo Collegato di Salerno, Via Ponte don Melillo, I-84084 Fisciano (SA),
Italy}
}

\begin{abstract}
We study state engineering through bilinear interactions between two remote qubits and two-mode Gaussian light fields. The attainable two-qubit states span the entire physically allowed region in the entanglement-versus-global-purity plane. Two-mode Gaussian states with maximal entanglement at fixed global and marginal entropies produce maximally entangled two-qubit states in the corresponding entropic diagram. 
We show that a small set of parameters characterizing extremally entangled two-mode Gaussian states is sufficient to control the engineering of extremally entangled two-qubit states, which can be realized in realistic matter-light scenarios.
\end{abstract}
\date{\today}
\pacs{03.67.Mn, 42.50.Dv, 42.50.Pq, 03.67.Hk}

\maketitle

The structure of quantum correlations within a given system depends strongly on the dimension of the state spaces of its constituents. The relation between correlations and system dimensionality becomes particularly relevant when considering states of bipartite systems whose parties are defined on Hilbert spaces of different dimension. An extreme instance is that of a bipartite compound consisting of a continuous variable (CV) system with infinite-dimensional state space and a discrete system with finite $D$-dimensional Hilbert space. This situation is particularly relevant in quantum communication, where 
many advantages come from the use of interfaces between light fields and matter-like systems~\cite{qinternet}, which are the basis of recent important experimental demonstrations~\cite{qinternet,sanders,referee1}. Ground-breaking architectures for communication, such as quantum repeaters~\cite{sanders}, rely on interfaces and their efficient implementation is based on the availability of such technological {\it primitives}. A {``desideratum"} of any reliable interface would be its ability to connect systems of different dimensionality so as to transfer important physical features (such as entanglement) across the interfaced systems while faithfully respecting their inherent {structure}. Among the proposals for quantum interfaces put forward so far~\cite{qinternet,referee1}, those aiming at transferring entanglement from a light field encoding a CV state to a qubit system~\cite{kraus,mauroprl2004} are appealing for implementations in a number of physical settings, ranging from cavity- and circuit-quantum electrodynamics (QED)~\cite{kraus,circuitQED} to polar molecules close to superconducting resonators~\cite{andre} or quantum dots and color-centers in diamonds in defect-microcavities and photonic crystals. Yet, a problem in dealing with mismatched system dimensionality is the loss of any clear relation between the purity and entanglement properties of the CV resource and those of an addressed two-qubit system. Controlling that the state hierarchy properties of the fields are inherited by the qubits is thus difficult.

{In this paper we discuss a simple and flexible CV-to-qubit map, based on the use of handy Gaussian-light resources, which works as a powerful tool for quantum state-engineering of the steady-state of two remote qubits. The rich variety of mapped two-qubit states faithfully respects the state hierarchy induced by the degrees of local/global purities and entanglement, whose structure is inherited from the Gaussian resources. 
Extremal bipartite Gaussian states (in terms of entanglement and mixedness) give rise to equally extremal two-qubit states. Our results thus shed new light  on the mechanisms at work at the interface between discrete and CV systems. Pragmatically, we show that our map represents an important step in the long-sought task of distributing channels for quantum communication~\cite{qinternet} and is implementable with resources available in many optical labs and relying on exisiting mature technology in cavity- and circuit-QED~\cite{circuitQED,georgiades}.}

\noindent {\it The map.}-- To keep the discussion as concrete as possible, in the following we will adopt the language of cavity QED, although our scheme is completely general and independent of any specific physical setting. Two remote single-mode cavities contain a qubit each and are coupled to a broad-band two-mode driving field prepared in an entangled state $\varrho_{12}$~\cite{kraus,mauroprl2004}. Qubit $A$ is coupled to cavity mode $1$ via the well-known Jaynes-Cummings Hamiltonian ${\hat{H}_{A1}\!=\!\omega_{} (\hat{q}_1\hat{\sigma}^1_A  + \hat{\sigma}^2_A \hat{p}_1)}$, where $\{\hat{q}_j ,\hat{p}_j\}$ are the quadratures of mode ${j=1,2}$ and $\hat{\sigma}_{k}^{1(2)}$ is the $x$ ($y$) Pauli matrix of qubit ${k\!=\!A,B}$. A similar model holds for qubit $B$ and mode $2$. Here, $\{\ket{0},\ket{1}\}_k$ are the logic states of qubit $k$ and $\omega$ is the Rabi frequency. The driven cavities leak photons at rate $\kappa$ and embody local environments for the qubits. We are interested in the control and manipulation of the entanglement and purity properties of the two-qubit system by means of its interaction with the bosonic field. States of CV systems can be described in terms of a (generally infinite) hierarchy of moments of their quadratures. Without loss of generality, the first moments are hereafter set to zero as they play no role in characterizing entanglement and mixedness. The second moments are used in the {\it covariance matrix} (CM) ${\sig}_{12}$ of elements ${{\rm Tr}[\varrho_{12} \{\hat x_\alpha, \hat x_\beta\}]}~({\alpha,\beta=1,..,4})$ with ${\hat{\bf x}\!=\!\{\hat q_1,\hat p_1, \hat q_2, \hat p_2\}}$. Gaussian states are completely specified by the knowledge of the first and second moments~\cite{Gaussiansomething}.
The CM of any two-mode state can be brought in the form
%\begin{equation}
%\label{cmsf}
${\bf V}_{12} =
\begin{pmatrix}
{\bf V}_1 & {\bf C}_{12} \\
{\bf C}^T_{12} & {\bf V}_2
\end{pmatrix}$
%\end{equation}
with ${\bf V}_1\!=\!\diag[a,a]$ (${\bf V}_2\!=\!\diag[b,b]$) and ${\bf C}_{12}\!=\!\diag[c_+,c_-]$.  In order for $\varrho_{12}$ to be a physical state, we must satisfy both $\sig_{12}\!>\!0$ and the uncertainty inequality $\sig_{12}{+}i\gr\Omega{\ge}0$ with $\gr\Omega\!=\!\bigoplus_{k=1,2}i\hat{\sigma}^{2}_{k}$ the symplectic matrix. 
In the bad-cavity limit where modes $1$ and $2$ reach their stationary state sooner than any changes in the qubit-field subsystems and within the first Born-Markov approximation, the state of qubits $A$ and $B$ evolves according to the master equation (ME)~\cite{mauroprl2004}
\begin{equation}
\label{liouv}
\partial_\tau \varrho_{AB}\!=\!\sum_{j,k=1}^4 \!d_{jk}(\hat{O}_j \varrho_{AB} \hat{O}_{k}\!-\!\{\hat{O}_k \hat{O}_j, \varrho_{AB}\}/2),
\end{equation}
where ${\hat O_j{=}\hat \sigma^j_A \otimes \id_B}$ for ${j{=}1,2}$ and ${\hat O_j{=}\id_A \otimes \hat \sigma^{j-2}_B}$ for $j{=}3,4$. The Kossakowski matrix ${\bf D}$ of elements $d_{jk}$ reads ${\bf D}{=}\gamma (\sig_{12}{+}i \gr\Omega)$, where $\gamma{=}2 \omega^2/\kappa$ is the effective qubit-field coupling strength, $\kappa$ is the cavity decay rate, and $\tau{=}\gamma t$ is the dimensionless time~\cite{notaSE}. The map in \eq{liouv} is completely positive iff $\bf D{\ge}0$~\cite{mauroprl2004}, which is equivalent to the uncertainty principle for the field CM. Thus the mapping holds for any (Gaussian or non-Gaussian) two-mode state with {\it bona fide} CM. Here, we consider only Gaussian states (generally mixed and asymmetric), bearing in mind that our results hold also for maps driven by non-Gaussian states with the same CM~\cite{notacm}. 

In Ref.~\cite{EPAPS} we study the dynamics of the two-qubit system. Here we deal with the features of $\varrho_{AB}(\infty)$, which is found by setting ${\partial_\tau \varrho_{AB}(\tau)\!=\!0}$ in the ME,  in relation to $\varrho_{12}$. By calling ${\varrho_{ij,kl}{=}\langle{ij}|\varrho_{AB}(\infty)|kl\rangle}$ (with $\ket{ij}$ and $\ket{kl}$ states of the two-qubit basis and ${i,j,k,l\!=\!0,1}$), we get ${\varrho_{00,00}\!=\![(ab\!-\!a\!-\!b)z\!+\!(a+b)^2]/\delta}$, $\varrho_{01,01}\!=\!\varrho_{00,00}+2[az-(a+b)^2]/\delta$, $\varrho_{10,10}\!=\!\varrho_{00,00}+2[bz-(a+b)^2]/\delta$,
${\varrho_{00,11}\!=\!{\varrho_{11,00}}\!=\!{2(a+b) (c_-\!-\!c_+)/\delta}}$, 
${\varrho_{01,10} \!=\!{\varrho_{10,01}}\!=\!2(a+b)(c_-\!+\!c_+)/\delta}$ with ${\delta\!=\!4abz}$, ${\varrho_{11,11}=1\!-\!\varrho_{00,00}\!-\!\varrho_{01,01}\!-\!\varrho_{10,10}}$, 
and ${z\!=\!(a+b)^2\!-\!2(c_+^2+c_-^2)}$. 

 \noindent {\it Degrees of entropy and entanglement}.-- The mixedness (or lack of purity) of the state $\varrho$ of a $D$-dimensional system can be quantified by the linear entropy ${S_L(\varrho) = [D/(D{-}1)] (1{-}{\rm Tr} \varrho^2)}$, ranging from $0$ (pure states) to $1$ (totally mixed states). For a CV Gaussian state $\varrho_{12}$ with CM $\sig_{12}$ one has ${S_L(\varrho_{12}){=}1{-}1/\sqrt{\det\sig_{12}}}$. Both for two-qubit and two-mode Gaussian states,
 separability is equivalent to positivity of the partially transposed density matrix~\cite{PPT}. The degree of violation of such a criterion provides an entanglement monotone, the {\it negativity} ${{\cal N}(\varrho_{AB})\!=\!\max\{0, \| \varrho_{AB}^{T_A}\|_1\!-\!1\}}$~\cite{werner}, where $T_A$ stands for partial transposition with respect to qubit $A$ and ${||\cdot||}_1$ is the trace-norm. For a two-mode Gaussian state $\varrho_{12}$ one has ${{\cal N}(\varrho_{12}){=}\max\{0,(1{-}\tilde{\nu}^-_{12})/\tilde{\nu}^-_{12}\}}$,
 where  $\tilde{\nu}^-_{12}$ is the smallest symplectic eigenvalue of the partially transposed CM, ${\tilde{\nu}^-_{12} =(1/\sqrt{2})[{\tilde{\Delta}_{12} - ({\tilde{\Delta}_{12}^2 - 4 \det \sig_{12}})^{\frac{1}{2}}}]^{\frac{1}{2}}}$, with ${\tilde{\Delta}_{12}\!=\!\det\sig_1\!+\!\det\sig_2\!-\!2\det{\bf C}_{12}}$. A two-mode Gaussian state %with CM $\sig_{12}$
is entangled if and only if ${\tilde{\nu}^-_{12}<1}$.
We parameterize $\sig_{12}$ by setting ${a=s+d,\,b=s-d}$ and
%\begin{equation}
%\label{param}
$c_{\pm}\!=\!\frac{\sqrt{(f_d-h_d)^2-4g^2}\!\pm\!\sqrt{(f_s-h_d)^2-4g^2}}{4 \sqrt{s^2-d^2}}$
%\end{equation}
with ${h_d\!=\!(2d^2\!+\!g)(\lambda\!+\!1)}$ and ${f_x\!=\!4 x^2\!+\!(g^2\!+\!1)(\lambda\!-\!1)/2}$ (${x\!=\!d,s}$)~\cite{ordering}. The Gaussian state $\varrho_{12}$ (with purity ${S_L(\varrho_{12})\!=\!1-g^{-1}}$) is physical {\it and} entangled for ${s \ge1,|d|\le s-1, 2|d|+1 \le g \le 2s-1}$ and ${\lambda{\in}[-1,1]}$, where $s$ and $d$ determine the properties of the reduced states $\varrho_{1(2)}$ having CM ${\sig}_{1(2)}$ according to ${S_L(\varrho_k){=}1{-}[s{+}(-1)^{k-1}d]^{-1}}$
with  $(k=1,2)$. At set values of $g$, $d$ and $s$, ${\cal N}(\varrho_{12})$ grows monotonically with $\lambda$.

What properties of the two-mode state are transferred to the two-qubit system? Marginal properties are faithfully reproduced as ${S_L(\varrho_{A(B)})=S_L(\varrho_{1(2)})[2-S_L(\varrho_{1(2)})]}$, 
where ${\varrho_{A(B)}}$
is the reduced state of qubit $A\,(B)$. The proportionality between $S\!_L(\varrho_{A(B)})$ and $S_L(\varrho_{1(2)})$ entails that the state-symmetry is preserved by the map. However, one finds that $S_L(\varrho_{AB})=1-1/(3 a^2)-1/(3b^2)-{16
 (a+b)^2}\xi^2/({3\delta^2})$ (with ${\xi^2=[(a+b)^2\!+\!4(c^2_++c^2_-)]}$), which shows that the two-qubit mixedness is not a simple function of $g$ alone,
but depends nontrivially also on $s, d$, and $\lambda$. In particular, %one can see that 
in the allowed range of parameters, $S_{L}(\varrho_{AB})$ increases with the global field mixedness ({\it i.e.} with $g$) and its mean energy  (parameterized by $s$), but decreases with $\lambda$: larger input entanglement at given entropy results in qubit states of higher purity. Notwithstanding the effective non-unitary dynamics, when $\varrho_{12}$ is pure (which occurs  when $a{=}b$ and ${c_+{=}-c_-{=}\sqrt{a^2-1}}$) $\varrho_{AB}(\infty)$ is also pure.
In general, for a given field mixedness, $S_{L}(\varrho_{AB})$ cannot vary unconstrained. Analytically, we find that $S_L(\varrho_{AB})$ admits tight upper and lower bounds that depend on the field mixedness $S_L(\varrho_{12})$. This is further confirmed by random numerical sampling [cfr.~Fig.~\ref{figrandom}{\bf (a)}]. The maximum of $S_L(\varrho_{AB})$ at a given $g$ is found by optimizing over $s,d$ and $\lambda$. This implies taking ${s{\gg}{1}
,\,\lambda{= -}1}$ and $d{=}0$. The corresponding two-qubit states tend asymptotically to the Werner state
 ${\varrho_{AB}^W =p \ket{\Phi^-}\!_{AB}\bra{\Phi^-} + {(1-p)}\id/4}$,
where $\ket{\Phi^-} _{AB}=(\ket{00}-\ket{11})_{AB}/{\sqrt{2}}$ and $p{=}2/(1+g^2)$, for which ${S_L^{\max}(\varrho_{AB})\!=\!1\!-\!4[1\!+\!(S_L(\varrho_{12})-1)^{-2}]^{-2}}$. On the other hand, $S^{\text{min}}_L(\varrho_{AB})$ at a given $g$ is obtained for $s=(g+1)/2$, $\lambda{=}1$ and $d=(g-1)/2$. Such  $S_L^{\min}(\varrho_{AB})$ is achieved by the product states $[{(g{-}1)}\ket{01}\!\bra{01}{+}{(g{+}1)}\ket{11}\!\bra{11}]/({2 g})$, for which we have ${S_L^{\min}(\varrho_{AB})=(2/3)S_L(\varrho_{12})[2-S_L(\varrho_{12})]}$. {\it The protocol can also be used to realize state purification: one finds many mapped $\varrho_{AB}$'s whose mixedness is smaller than the input $S_L(\varrho_{12})$, even for totally mixed fields} [see Fig.~\ref{figrandom}{\bf (a)}]. Noticeably,
two-qubit states obtained from highly mixed $\varrho_{12}$'s are separable while entanglement arises in the region of moderate mixednesses.
\begin{figure}[t]
\hskip0.5cm{{\bf (a)}\hskip3.0cm{\bf (b)}}\\
%\subfigure[]
{\includegraphics[width=2.9cm]{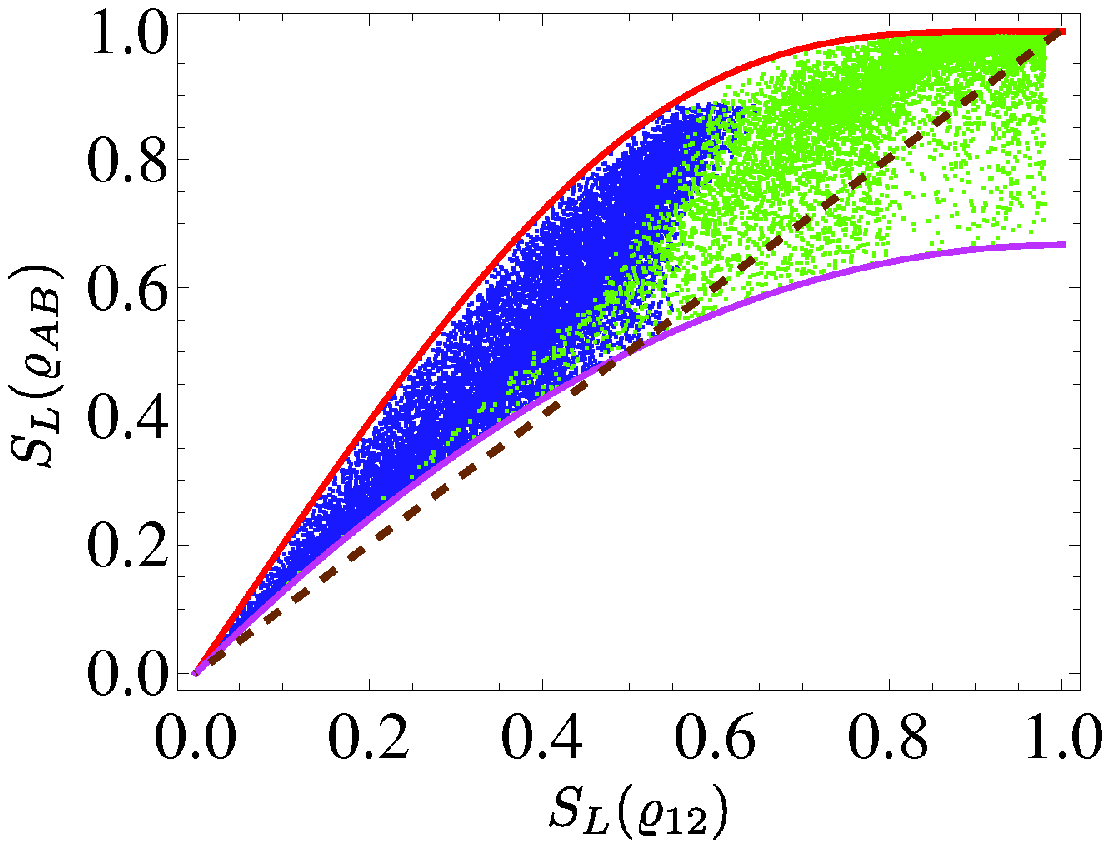}} \hspace{.3cm}
%\subfigure[]
{\includegraphics[width=2.9cm]{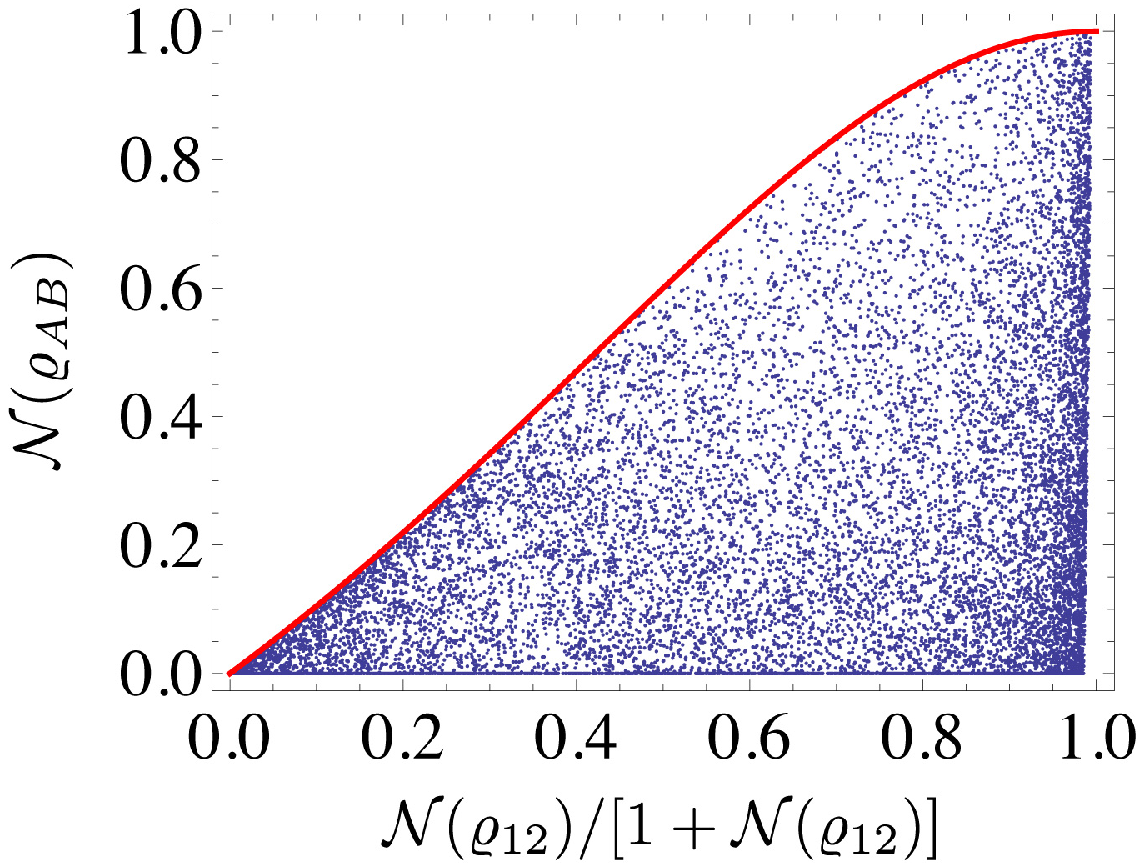}}
\caption{(Color online). {\bf (a)} $S_L(\varrho_{AB})$ versus $S_L(\varrho_{12})$ for $20000$ random states. Darker (lighter) dots denote entangled (separable) qubit states. Points below the dashed line 
show states purified by our map.
{\bf (b)} ${\cal N}(\varrho_{AB})$ versus normalized field negativity for $15000$ random states. 
\label{figrandom} }
%}
\end{figure}
By introducing  ${\eta{=}\frac{2(a+b)(c_+\!-\!c_-)}{z(a\!-\!b)}}$ and $\mu=z(a-b)(1+\eta^2)^{1/2}$, the two-qubit negativity at the steady-state reads
\begin{equation}
\label{nab}
%\begin{aligned}
{\cal N}(\varrho_{AB}){=}\max\{0,({2}/{\delta})[(a+b)^2\!-\!{\delta}/{4}\!+\!\mu]\}.
\end{equation}
In the physically allowed range of parameters, we have ${\partial_g {\cal N}(\varrho_{AB})\!\le\!0}$, ${\partial_{s,\lambda}{\cal N}(\varrho_{AB})\!\ge\!0}$
while the dependence on $d$ is non-monotonic. This implies that the two-qubit entanglement increases with the marginal entropies, decreases with the global mixedness and, at fixed global and marginal entropies, increases with $\lambda$. {\it These are the very same patterns followed by ${\cal N}(\varrho_{12})$ and therefore the map in Eq.~(\ref{liouv}) fully preserves the qualitative structure of bipartite entanglement}. Quantitatively, however, ${\cal N}(\varrho_{12})$ does not determine directly the  two-qubit negativity. For a given value of ${\cal N}(\varrho_{12})$, the corresponding $\varrho_{AB}$ range from separable to highly entangled. The behavior of ${\cal N} (\varrho_{AB})$ versus the negativity of randomly sampled CMs ${\bf V}_{12}$ is reported in Fig.~\ref{figrandom}{\bf (b)}. While  ${\cal N} (\varrho_{AB})$ can vanish for arbitrarily entangled $\varrho_{12}$'s, we find a maximum of the two-qubit entanglement  ${\cal N}^{\max}(\varrho_{AB})\!=\!1\!-\!2[(1+{\cal N}(\varrho_{12}))^2\!+\!1]^{-1}$ that is achieved by pure states (both for fields and for qubits).

\noindent {\it Entanglement versus global mixedness.}-- Maximally entangled two-qubit mixed states (MEMS) are defined as those maximizing a given entanglement measure at any fixed value of the global mixedness~\cite{munro}. In the ${\{S\!_L(\varrho_{AB}),\,{\cal N}(\varrho_{AB})\}}$ space, MEMS include the family of Werner states. The corresponding minimally entangled mixed states are just separable states. The Gaussian counterparts to MEMS (GMEMS) are two-mode mixed states with infinite entanglement, such as the two-mode squeezed thermal states, with
 ${a\!=\!b\!=\!\sqrt{g} \cosh(2r)}$, ${c_+\!=\!-c_-\!=\!\sqrt{g} \sinh(2r)}$
 %\end{equation}
 in the limit ${r\rightarrow{\infty}}$. We remark that, given all possible CMs, our scheme
 {does not} generate  {\it every} possible two-qubit state. However, the set of $\varrho_{AB}$'s that can be engineered by our process {\it does} fill the entire region of the ${\{S_L(\varrho_{AB}),\,{\cal N}(\varrho_{AB})\}}$ diagram physically allowed to two-qubit states. This result is illustrated in Fig.~\ref{figmmmm}{\bf (a)} where we report the diagnostics of $\varrho_{AB}$'s obtained from random $\varrho_{12}$'s. The upper bound to the physically allowed region includes states $\varrho^W_{AB}$ for which ${\cal N}^{\max}(\varrho_{AB}){=}[-1{+}3 \sqrt{1-S_L(\varrho_{AB})}]/2$ if $S_L(\varrho_{AB}){<} 8/9$, and zero otherwise. A direct way to obtain such states is by maximizing $\lambda$, minimizing $d$ and setting ${g=1/p}$, which fixes the global purity of the two-mode resource to be equal to the $\ket{\Phi^-}$ component of  $\varrho^W_{AB}$. Interestingly, the  field state associated to such parameters is precisely a GMEMS
with ${g\!=\!1/p}$. The mapped two-qubit state converges to the corresponding boundary state when the squeezing in $\varrho_{12}$ is large. We can thus engineer MEMS of tunable entanglement/purity by adjusting purity and squeezing in ${\sig_{12}}$. In Ref.~\cite{EPAPS} we present an additional study of  ${\cal N}(\varrho_{AB})$ against the marginal entropies of the resource.

\begin{figure}[t]
\hskip-0.3cm{{\bf (a)}\hskip2.0cm{\bf (b)}}\hskip2.5cm{\bf (c)}\\
% \subfigure[]
 {\includegraphics[width=2.7cm]{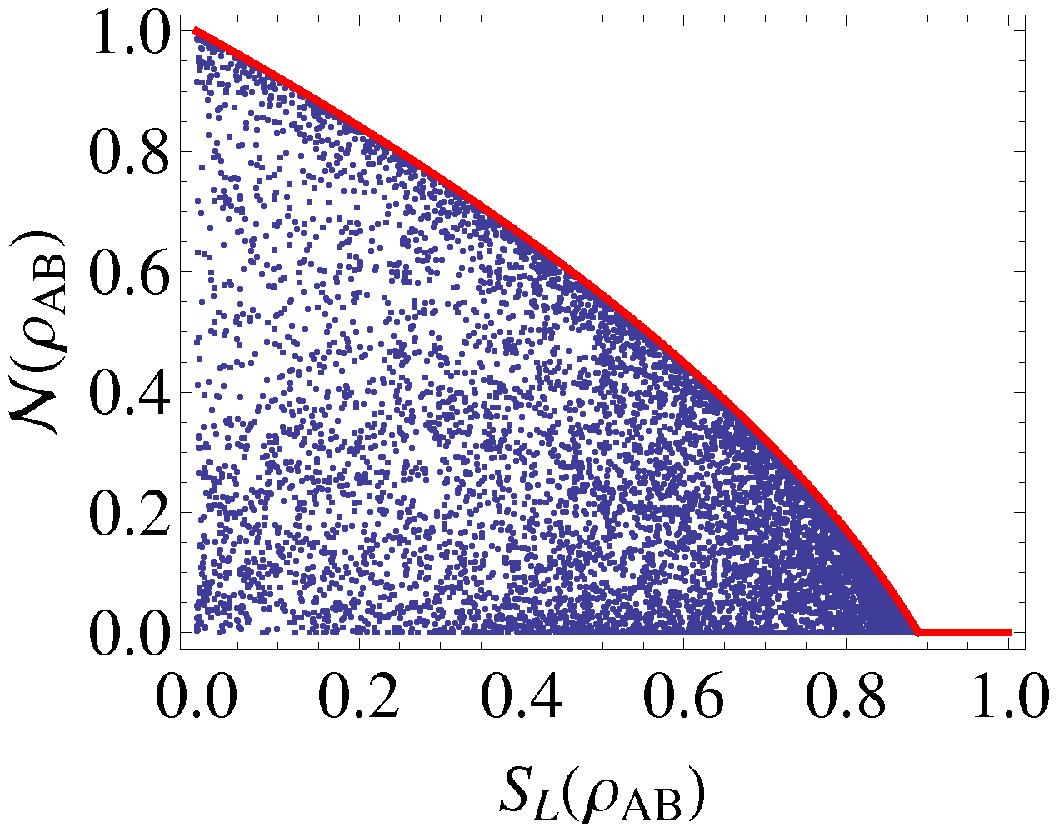}}
 {\includegraphics[width=2.9cm]{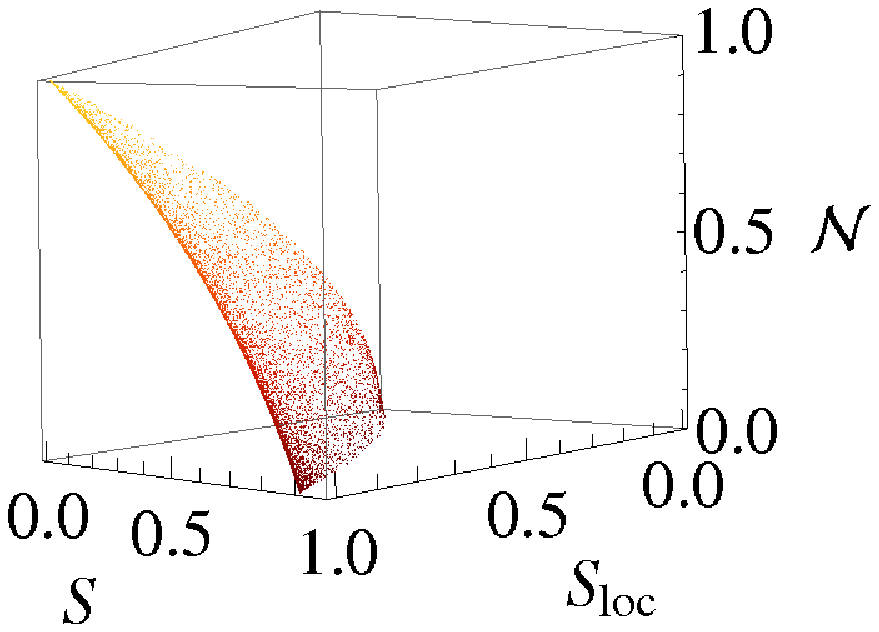}}
%\subfigure[]
 {\includegraphics[width=2.9cm]{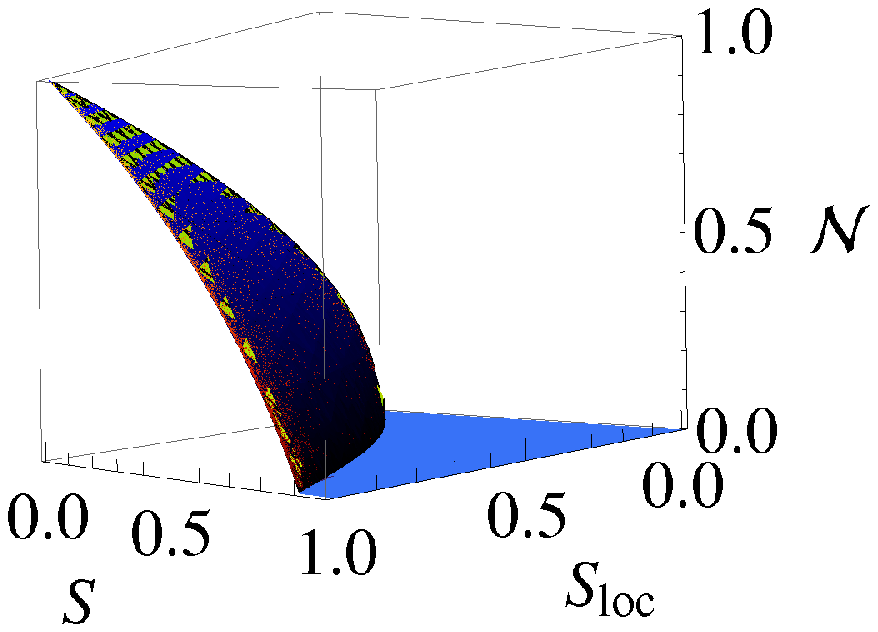}}
%\subfigure[]
\caption{(Color online). {\bf (a)}  ${\cal N}(\varrho_{AB})$ versus $S_L(\varrho_{AB})$ of two-qubit states obtained from $20000$ random entangled two-mode Gaussian states $\varrho_{12}$'s. The upper boundary (MEMS) includes Werner states. {\bf (b)} ${\cal N}(\varrho_{AB})$ versus $S_{\text{loc}}$ and $S$ for states obtained using $10^4$ random symmetric entangled $\varrho_{12}$'s. {\bf (c)} The same states as in {\bf (b)} and the surfaces of maximum and minimum negativity at fixed entropies.
}
\label{figmmmm}
%}
\end{figure}
\noindent 
{\it Entanglement versus global and marginal entropies.}--
 A very refined characterization of entanglement is possible in the space of global {\it and} marginal entropies, where all the entangled two-mode Gaussian states lie in a narrow region bounded by Gaussian least-entangled and Gaussian maximally-entangled mixed states (GLEMS and GMEMS, respectively)~\cite{extremal}. These are achieved, at given $s,d$, and $g$ (fixing the entropies) for ${\lambda\!=\!-1}$ (GLEMS) and ${\lambda\!=\!1}$ (GMEMS), respectively. Thus, by accessing only the restricted set of parameters that determine the marginal and global purities of a two-mode Gaussian state, one can pin-down its entanglement and bound it with the corresponding GLEMS and GMEMS.
 Surprisingly, no exact two-qubit counterpart to this structure was known thus far. We will now show that, via Eq.~(\ref{liouv}), the two-qubit states $\varrho_{AB}$ inherit and enhance the properties of
 such CV states.
 As ${\lambda \in [-1,1]}$, the existence of GLEMS and GMEMS stems from observing that ${\partial_\lambda
 {\cal N}(\varrho_{12})|_{s,d,g} \ge 0}$. A similar property holds for $\varrho_{AB}$. For ease of notation, we write ${S{\equiv}{S_L(\varrho_{AB})}}$ and ${{\cal N}_q{\equiv}{\cal N}(\varrho_{AB})}$.
From Eqs.~{(\ref{nab})} it follows that  ${\cal N}_{q}$ is a function of $s,d,\lambda$ and $g(s,d,\lambda,S)$ such that ${\partial_\lambda{\cal N}_{q}}{|}_{s,d,S}={\partial_\lambda{\cal N}_{q}}|_{s,d,g} +{\partial_{g}{\cal N}_{q}}|_{s,d,\lambda}{\partial_\lambda{g}}|_{s,d,S}={{\partial_\lambda{\cal N}_{q}}|_{s,d,g}}+{{\partial_g{\cal N}_{q}}|_{s,d,\lambda}}{({{\partial_\lambda{S}|_{s,d,g}/\partial_gS|_{s,d,\lambda}}}})\ge{0}$.
This shows the existence of two-qubit least-entangled and maximally-entangled mixed states (QLEMS and QMEMS, respectively), at fixed global and marginal entropies. These are obtained by mapping of GLEMS and GMEMS, respectively. The values of $s$ and $d$ are set by ${S\!_L(\varrho_{k})~(k=A,B)}$, while $g$ is determined by the marginal and  global entropies. For symmetric QMEMS with ${\lambda\!=\!1}$, ${S_L(\varrho_{k})\!\equiv\!S_{\text{loc}}\!=\!1{-}s^{-2}}$ and global mixedness $S$, one has ${g=3/[1\!-\!S_{\text{loc}}\!+\!\sqrt{4\!-\!9 S\!+\!S_{\text{loc}}\left(4\!+\!S_{\text{loc}}\right)}]}$. The corresponding ${\cal N}_{q}^{\max}$ (see Ref.~\cite{EPAPS}) fixes the upper bound for all two-qubit states obtainable by our process and compatible with the given entropies. An analogous analysis holds for QLEMS.

The behavior of the negativity versus global and marginal entropies shows that all the mapped two-qubit states lie in a quasi-bidimensional region [see Fig.~\ref{figmmmm}{\bf (b)}]. Even for mixed states, ${\cal N}_{AB}$ is almost perfectly a function of the global and marginal entropies alone. By superimposing the boundary curves corresponding to QLEMS and QMEMS to the numerical analysis in Fig.~\ref{figmmmm}{\bf (b)} we see that all the randomly generated states $\varrho_{AB}$'s accumulate in the tight interval between maximum and minimum negativity  [cfr. Fig.~\ref{figmmmm}{\bf (c)}]. Numerically, the negativities of QLEMS and QMEMS differ by less than $0.04$ e-bits close to the separability point, while in the region of larger entanglement they practically coincide. Therefore the field-qubit interface defines, within the set of entangled two-qubit states, the analogues to two-mode GMEMS and GLEMS. Moreover, the situation typical of the Gaussian case is enhanced in the case of qubits, since in the latter the gap between maximal and minimal entanglement is even narrower. The introduced mode-qubit dynamical interface preserves the hierarchy of entangled states in the entropic space: GMEMS (GLEMS) are mapped into QMEMS (QLEMS). Furthermore,  $\varrho_{AB}$ weakly depends  on $\lambda$ so that ${\cal N}(\varrho_{AB})$ can be  accurately determined only controlling the engineered entropies of $\varrho_{12}$.\\
\noindent
{{\it Practical considerations.--} Cavity QED is a natural setting for the implementation of the proposed scheme, due to the availability of a variety of CV resources~\cite{ralph}. 
The feeding of an optical cavity containing a trapped atom with squeezed light (bandwith of about $12$ MHz) and the corresponding controlled light-atom interaction have been experimentally demonstrated~\cite{georgiades} for ${(\kappa,\omega)/2\pi\simeq({70},20)}$ MHz. This gives an effective coupling rate ${\gamma\simeq{10}}$ MHz. The engineering of MEMS from approximated GMEMS produced at high squeezing is thus feasible. The swift progress in circuit-QED makes also such setting appealing~\cite{circuitQED}. A qubit is embodied by a superconducting quantum interference device (SQUID) at the charge-degeneracy point. The qubit transition-energy can be set by an \emph{in situ} magnetic flux that modulates the Josephson energy so as to adjust the qubit-to-light coupling. Each qubit is integrated in a  full/half-wave waveguide split by input/output capacitances: we can thus consider two disconnected regions of a coplanar waveguide, joined via independent input/output capacitive lines for the injection/leakage of the field resource or via a large Josephson junction. The resonator quality-factor is typically well within a range appropriate to our scheme ($10^2$ to $10^6$ with $\omega{\sim}0.1$ GHz). For the frequencies involved in this setup, this would allow the realization of the regime studied here. Very recently, the ability to perform complete state tomography of two such qubits has been experimentally demonstrated~\cite{tomo}, opening up the possibility for preparation and characterization of the qubit state. Although the ability to experimentally engineer microwave-field states are inferior to their optical counterparts, squeezed-vacuum/thermal states have been produced by Josephson parameteric amplifiers embodied by large junctions~\cite{lenhert}, which is a very promising step towards the preparation of microwave states belonging to the classes studied here. Our scheme may be used to entangle collections of remote matter-like qubits, so as to achieve key resources in quantum technology.}\\
\noindent
{\it Conclusions.--} We have discussed a CV-to-qubit map that engineers two-qubit states spanning the region of physically allowed quantum correlations at fixed values of local and global entropies. It incorporates the most relevant sources of noise affecting a CV-to-qubit interface. Our results assure
the realistic possibility for non-demanding production and control of qubit states using off-line preparation of CV entangled resources and linear local interactions. The relations found between two-mode fields and two-qubit states make our scheme a basic predictive tool for light-matter entanglement transfer and related implementations in quantum technology.\\
\noindent
{\it Acknowledgments.--} We thank DEL, EPSRC, the FP7 project HIP, MIUR under the FARB Fund, INFN under Iniziativa Specifica PG62, CNR-SPIN and the ISI Foundation.

\clearpage

Supplementary Material

\maketitle

\renewcommand{\theequation}{S-\arabic{equation}}
\setcounter{equation}{0}  % reset counter 

\section{Quantitative analysis of the dynamical CV-to-qubit map}

In this Section we provide further details on the map embodied by the master equation (ME) in Eq.~(2) of the main paper. Rather than discussing the steady state properties of the mapped two-qubit state, here we address the full dynamical evolution of an initial preparation of qubits $A$ and $B$. 

When written in the computational basis $\{\ket{00},\ket{01},\ket{10},\ket{11}\}_{AB}$, $\varrho_{AB}$ can be partitioned as $\varrho_{AB}(\tau)=\varrho_{AB}^{\mbox{\footnotesize \sf x}}(\tau)+ \varrho_{AB}^{\mbox{\footnotesize \sf o}}(\tau)$, where 
\begin{equation}
\varrho_{AB}^{\mbox{\footnotesize \sf x}}(\tau)= 
\begin{pmatrix}
\star  &   &   & \star  \\
   & \star  & \star  &   \\
   & \star  & \star  &   \\
 \star  &   &   & \star
\end{pmatrix},
~
\varrho_{AB}^{\mbox{\footnotesize \sf o}}(\tau)=
\begin{pmatrix}  
 & \star  & \star  &   \\
 \star  &   &   & \star  \\
 \star  &   &   & \star  \\
   & \star  & \star   
\end{pmatrix}.
\end{equation}
Here, the symbol $\star$ is used to denote any potentially non-zero matrix entry. Obviously, $\varrho_{AB}^{\mbox{\footnotesize \sf o}}(\tau)$ is not a density matrix. 
It is straightforward to check that the Liouvillian $\hat{\cal L}$ in Eq.~(2) of the main paper keeps $\varrho_{AB}^{\mbox{\footnotesize \sf x}}(\tau)$ disjoint from $\varrho_{AB}^{\mbox{\footnotesize \sf o}}(\tau)$, so that starting from a state having elements only in $\varrho_{AB}^{\mbox{\footnotesize \sf x}}(\tau)$, we are sure that $\varrho_{AB}^{\mbox{\footnotesize \sf o}}(\tau)=0~\forall{\tau}$. This is clearly the case for $\varrho_{AB}(0)\!=\!\ket{00}_{AB}\bra{00}$, and this observation yields a great simplification enabling us to swiftly determine the dynamics of the entanglement transferred to the qubits. Moreover, such an initial condition is the most favorable one to achieve the highest possible qubit entanglement in the model under investigation.

\begin{figure}[b]
\center{{\bf (a)}}\\
\center{\includegraphics[width=7cm]{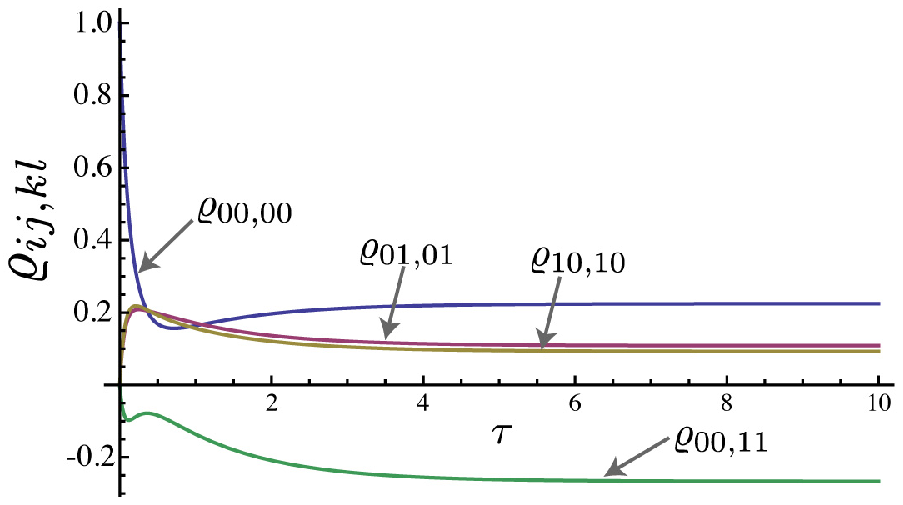}}
\center{{\bf (b)}}\\
\center{\includegraphics[width=7cm]{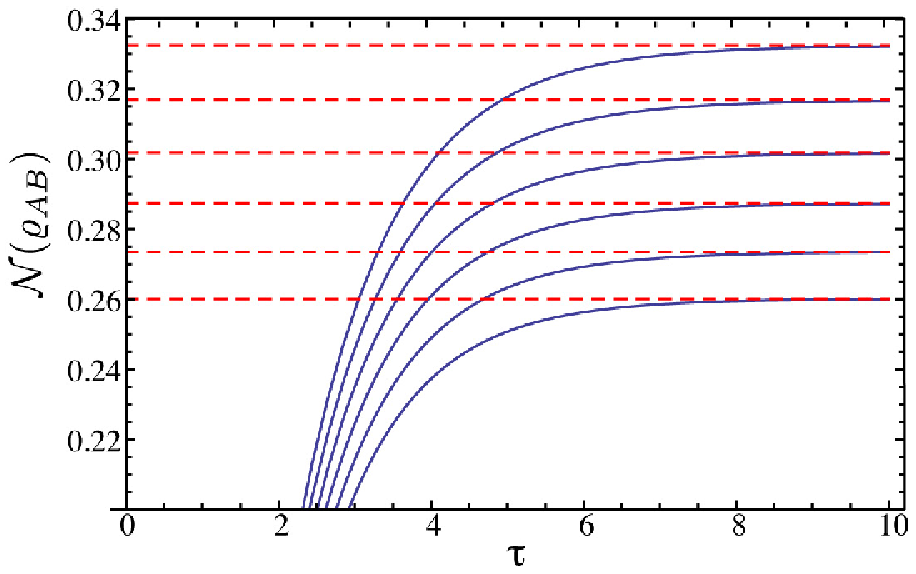}}
\caption{(Color online). {\bf (a)}  Behavior of the density matrix elements against the dimensionless interaction time $\tau$ for (randomly taken) $s=1.774,\,d=0.07,\,g=1.448,\gamma=0.1$ and $\lambda=1$. {\bf (b)} Dynamical behavior of the negativity ${\cal N}(\varrho_{AB})$ against $\tau$ for the same values as in panel {\bf (a)} but $\lambda$, which is taken to grow from $-1$ to $1$ in steps of $0.4$. The lower-most (top-most) curve is for $\lambda=-1$ ($\lambda=1$). The dashed horizontal lines show the steady-state values.
} 
\label{figdyn} 
%}
\end{figure}

Starting from the ME in Eq.~(2) of the manuscript, it is straightforward to write down the Bloch-like equations for the evolution of the two-qubit density matrix elements and solve them against the dimensionless interaction time $\tau$ for any specific assignment of the resource's covariance matrix (CM). Fig.~\ref{figdyn}{\bf (a)} shows the typical $\tau$-dependent behavior of such elements, where we have used the notation ${\varrho_{ij,kl}=\langle{ij}|\varrho_{AB}(\tau)|kl\rangle}$ (with $\ket{ij}$ and $\ket{kl}$ states of the two-qubit basis and ${i,j,k,l\!=\!0,1}$), similarly to what has been done in the main paper. Similar trends are found for any allowed choice of parameters in $\sig_{12}$. Interesting information can be extracted from the dynamical behavior of $\varrho_{AB}(\tau)$. First, by fixing $s,d$ and $g$ in $\sig_{12}$, one can easily see that the dynamical evolution of the negativity  of the two-qubit system keeps ${\cal N}(\varrho_{AB})$ within a very narrow range of values, as $\lambda$ is varied (we remind that $\lambda\in[-1,1]$). This is in line with the behavior highlighted in Fig.~2{\bf (b)} and {\bf (c)} of the main paper, although here we are not imposing any restriction to the local/global entropies of the Gaussian resource state. Fig.~\ref{figdyn}{\bf (b)} shows, in this sense, an {\it unusual scenario} where the difference between the negativity at $\lambda=-1$ and the corresponding value at $\lambda=1$ isabout $30\%$. Typically the top-most curve would be less than $5\%$ far from the bottom one. 

\begin{figure}[t]
{\includegraphics[width=7cm]{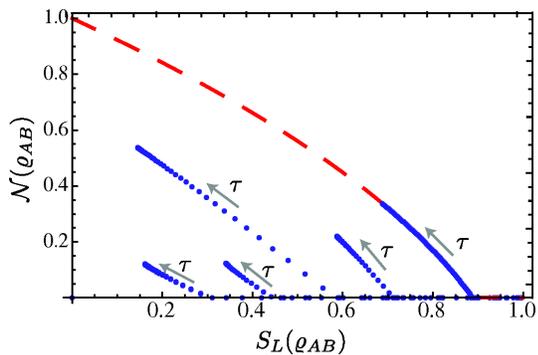}}
\caption{(Color online). Typical trajectories of the mapped two-qubit states $\varrho_{AB}$ in the $\{{S}(\varrho_{AB}),{\cal N}_L(\varrho_{AB})\}$ plane. The dashed boundary curve embodies maximally entangled mixed states (MEMS) and Werner states $\varrho^W_{AB}$. The dots show the {\it evolution} of negativity and global linear entropy of the two-qubit state for five different choices of the Gaussian resource parameters $s,g,d$ and $\lambda$. The dimensionless interaction time $\tau$ grows as indicated by the arrow. The final dot in each trajectory indicates the corresponding steady state. The trajectory superimposed to the MEMS curve is for a configuration of the parameters entering $\sig_{12}$ which guarantees the asymptotic mapping to a $\varrho^W_{AB}$ state.} 
\label{figMEMS} 
\end{figure}

\begin{figure}[b]
{\includegraphics[width=7cm]{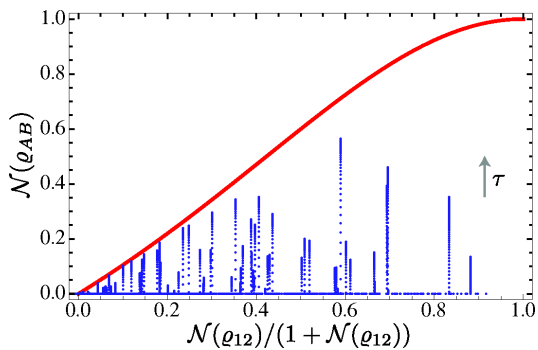}}
\caption{(Color online). Typical trajectories of the mapped two-qubit states $\varrho_{AB}$ in the plane spanned by the (normalized) negativity of the Gaussian resource ${\cal N}(\varrho_{12})$ and the target qubit-state negativity ${\cal N}(\varrho_{AB})$. The boundary curve follows the functional form found in the main paper. The dots along each vertical line show the {\it evolution} of the two-qubit negativity for random choices of the Gaussian resource parameters $s,g,d$ and $\lambda$. The dimensionless interaction time $\tau$ grows as indicated by the arrow (at $\tau=0$, obviously, each two-qubit state is separable). The final dot in each trajectory indicates the corresponding steady state.}
\label{figInput} 
\end{figure}

The information gathered throughout the dynamical evolution of the qubit density matrix allows one to infer how the two-qubit system approaches its steady state, for a given choice of the input CM. By using $\tau$ as a sort of {\it curvilinear abscissa} of the negativity-versus-linear entropy functions, we can determine the trajectories of the two-qubit density matrices up to the boundary curve accommodating maximally entangled mixed states and Werner states [see Fig.~\ref{figMEMS}]. A similar study can be conducted in order to infer the dynamical mapped-state entanglement against the resource negativity (as in Fig.~\ref{figInput}). We refer to the captions of such figures for further details on the simulations. 

\section {Entanglement versus marginal mixednesses.}

\begin{figure}[t]
{\includegraphics[width=7cm]{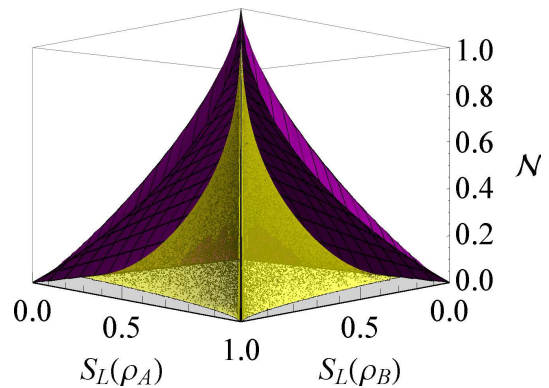}}
\caption{(Color online). Two-qubit negativity against local marginal entropies. Physical two-qubit states lie below the top tent-like boundary (MEMMS). Such states are not reproducible using Gaussian states as these produce $\varrho_{AB}$'s lying in the inner region, whose upper boundary contains states engineered via GMEMMS. \label{figmmmm1} }
%}
\end{figure}

Here we classify the bipartite entanglement in the marginal-entropy space. Such a characterization is exact for pure states. In general, bounds on mixed-state entanglement can be derived for all mixed states compatible with given marginals. For two qubits, the upper bound to the physical set of states in this diagram defines {\it maximally entangled states at fixed marginal mixednesses} (MEMMS)~\cite{memms}, while separable (product) states exist for any pair of marginals~\cite{footnote}. For equal marginal entropies, MEMMS are pure states. Their Gaussian counterparts have been introduced in Ref.~\cite{extremal} and dubbed Gaussian-MEMMS, or GMEMMS.
Their CM is characterized by having ${g=2|d|\!+\!1}$. Correspondingly, any dependence of the CM on $\lambda$ disappears and one simply has ${c_\pm = \pm  \sqrt{(1+\max\{a,b\}) (-1+\min\{a,b\})}}$~\cite{extremal}. 

A random-state investigation of the performances of our map in such local-entropy space shows that the mapped two-qubit steady states do not fill the whole region allowed to any two-qubit state with given marginals. In particular, any attempt to reproduce MEMMS results in unphysical parameters for the driving field. In Fig.~\ref{figmmmm1} we show such a numerical exploration of the negativity ${\cal N}(\varrho_{AB})$ against the marginal entropies $S_L(\varrho_{A})$ and $S_L(\varrho_{B})$. The attainable qubit state fill a restricted, pyramid-like region: their entanglement is never maximal at given marginals, except in the pure-state case of $S_L(\varrho_{A})=S_L(\varrho_{B})$. We now aim at characterizing analytically the upper boundary of this set. Recalling that the negativity is an increasing function of $\lambda$ and a decreasing function of $g$ (at fixed $a$ and $b$), we can conclude that the boundary is obtained by minimizing $g$, which gives $g=2|d|+1$. Such values of the parameters are exactly those characterizing a GMEMMS. Although the full range of entanglement is not achievable in the space of the marginal mixedness, our map is such that the two-qubit states endowed with the maximum achievable entanglement at fixed marginals are those obtained by using GMEMS. Therefore, despite being unable to reproduce MEMMS, the maximum negativity achieved by our scheme corresponds to the images of GMEMMS. Fig.~\ref{figmmmm1} provides further details on this point. We believe this is yet another clear indication of the powerful and faithful mapping embodied by by our simple bilocal linear interaction model.

\section{Analytic expression for the negativity of QMEMS}

We consider the space of negativity against global and marginal mixedness. The aim of this Section is to give an explicit form to the negativity of two-qubit most-entangled mixed states (QMEMS). By taking the case of a resource embodied by a Gaussian most entangled mixed states (or GMEMS) with $\lambda=1$ and imposing the symmetry conditions given by $S_L(\varrho_A)=S_L(\varrho_B)=1-1/s$ (we call $S_L(\varrho_{AB}){\equiv}{S}$ the global linear entropy), it is straightforward to check that the negativity of the extremal two-qubit states in this space  reads
\begin{equation}
\begin{aligned}
&{\cal N}_{q}^{\max}=\frac{1}{6}\left[-(2+S_{\text{loc}})+\sqrt{\left(2+S_{\text{loc}}\right)^2-9 S}\right.\\
&\left.+2\left(2+8S_{\text{loc}}-S^2_{\text{loc}}\!-\!9S+(S_{\text{loc}}\!-\!1)\sqrt{\left(2+S_{\text{loc}}\right)^2\!-\!9S}\right)^{\frac{1}{2}}\right]
\end{aligned}
\end{equation}
 {if} $9S+4(-2+S_{\text{loc}})S_{\text{loc}}\!<\!4S_{\text{loc}}\sqrt{1-S_{\text{loc}}+S^2_{\text{loc}}}$ and ${\cal N}_{q}^{\max}=0$ {otherwise}.  More cumbersome expressions are found for $g$ and the corresponding minimal value ${\cal N}_{q}^{\min}$ valid for two-qubit least entangled mixed states (QLEMS).


\begin{thebibliography}{99}
\bibitem{notacm} The entanglement passed by non-Gaussian states would not exceed that achieved with Gaussian ones, although the former are generally more entangled than the latter with the same CM~\cite{Gaussiansomething}.
\end{thebibliography}
\begin{thebibliography}{99}
\bibitem{footnote} In Ref.~\cite{memms} MEMMS have been characterized using concurrence. However, it is straightforward to adapt the formalism to negativity. One finds that that concurrence-based MEMMS also maximize negativity at fixed marginals.
\end{thebibliography}
\end{document}